\documentclass[preprint,amssymb,showpacs,prl,aps]{revtex4}
\usepackage{graphicx}
\begin{document}

\title{
       High frequency acoustic modes in liquid gallium \\ at the melting point.
      }
\author{
        T.~Scopigno$^{1}$,
        A.~Filipponi$^{2}$,
        M.~Krisch$^{3}$
        G.~Monaco$^{3}$
        G.~Ruocco$^{1}$,
        F.~Sette$^{3}$
        }
\affiliation{
    $^{1}$Dipartimento di Fisica and INFM, Universit\'a di Roma ``La Sapienza'',
I-00185, Roma, Italy.\\
    $^{2}$Dipartimento di Fisica and INFM, Universit\'a di L'Aquila, I-67010,
L'Aquila, Italy. \\
    $^{3}$European Synchrotron Radiation Facility, B.P. 220 F-38043 Grenoble,
Cedex France.
    }
\date{August 14, 2002}

\begin{abstract}

The microscopic dynamics in liquid gallium (l-Ga) at melting
(T=315 K) has been studied by inelastic x-ray scattering. We
demonstrate the existence of collective acoustic-like modes up to
wave-vectors above one half of the first maximum of the static
structure factor, at variance with earlier results from inelastic
neutron scattering data [F.J. Bermejo et al. Phys. Rev. E {\bf
49}, 3133 (1994)]. Despite the structural (an extremely rich
polymorphism and rather complex phase diagram) and electronic
(mixed valence) peculiarity of l-Ga, its collective dynamics is
strikingly similar to the one of Van der Walls and alkali metals
liquids. This result speaks in favor of the universality of the
short time dynamics in monatomic liquids rather than of
system-specific dynamics.

\end{abstract}


\maketitle


The nature of the microscopic dynamics in simple fluids is
nowadays one of the most lively debated topics in the condensed
matter field. The provocative evidence of well defined collective
excitations outside the truly hydrodynamic region has attracted
many experimentalists in the last three decades. In particular,
several studies performed through Inelastic Neutron Scattering
(INS) revealed the presence of a Brillouin triplet in the dynamic
structure factor of many monatomic liquids \cite{neu}. Together
with numerical studies \cite{num} these experiments provided us
some pictures of the high frequency dynamics, although a deep
comprehension of the ultimate nature of these excitations is
still missing \cite{books,teo}. More recently, the advent of the
new radiation sources has allowed the full development of the
Inelastic X-ray Scattering (IXS) \cite{ixs}. This technique,
being sensitive to the coherent dynamics only and allowing to
investigate the low exchanged momentum ($Q$) region inaccessible
to INS, renewed the interest and the efforts in the field giving
rise to a number of experiments \cite{ixsexp}. More specifically,
some light has been shed on the mesoscopic dynamics of different
monatomic systems as Van der Walls fluids (He, Ne), liquid alkali
metals (Li, Na), and more complex liquid metals (Al). In
particular, using the generalized Langevin equation formalism,
the presence of two viscous relaxation processes has been
experimentally proven in all these systems, and the relevant
parameters have been quantitatively determined \cite{scopixs}.

Among the elemental liquid metals, Ga exhibits peculiar
structural and electronic properties. In addition to the low
melting point ($T_m=303$ K), it presents an extended solid phase
polimorphism with complex crystal structures where a competition
between metallic and covalent bonding character takes place
\cite{tos}. Although the electronic DOS in l-Ga is nearly free
electron it still shows anomalies associated with some covalency
residue. Moreover, the first peak of the $S(q)$ presents a hump
characteristics of non close-packed liquid structures \cite{bf}.
The existence of common features in the high frequency dynamics
of all the monatomic liquids seems to be challenged by liquid
Gallium. In fact, earlier studies performed by INS showed somehow
contradictory results. At room temperature, although no
collective modes were visible in the experimental data, the
spectra were described by a damped harmonic oscillator model, and
the excitation frequencies were found to lie above the
hydrodynamic values. This effect was ascribed to the presence of
high frequency optical modes that were supposed to contribute to
the spectra. In these studies, the absence of acoustic
excitations was explained with a high value of the longitudinal
viscosity, despite on the basis of hydrodynamic arguments
collective modes should have been expected \cite{ber1}. Few years
later, a new INS experiment was performed at higher temperature
(T=973 K), and collective modes were detected. In particular, at
low wavevectors an acoustic mode was found, while at high
wave-vectors the dispersion curve split in two branches that
according to the authors should be associated to acoustic and
optic excitations respectively \cite{ber2}. This picture (absence
of acoustic-like excitations just above melting) -if verified-
would stress the anomalous dynamical behaviour of Gallium, and
would falsify the idea of a common high frequency dynamics of
monatomic systems.

In this paper we report on an inelastic x-ray scattering study of the
microscopic dynamics in liquid gallium just above the melting point. The
sensitivity of this technique to the purely collective motion only and the
extended accessible cinematic region, allowed an accurate determination of the
coherent dynamic structure factor. We find clear indication of acoustic-like
excitations, whose properties parallel those of all the other investigated
monatomic liquids. Moreover, the value of the transport coefficients derived
by a generalized Langevin equation description of the spectra, quantitatively
agrees with the hydrodynamic expectations. The obtained results indicate that
-despite its structural and electronic anomalies- liquid Gallium shares with
the other simple liquids (metals and non-metals) the same features of the high
frequency atomic dynamics.

The experiment has been performed at the ID28 beam-line of the
ESRF at fixed exchanged wave-vector over a $Q-$ region below the
position of the main diffraction peak ($Q_M \approx 25$
nm$^{-1}$). A typical energy scan ($-50 < E < 50$ meV) took about
300 minutes, and was repeated for a total integration time of
about 300 seconds/point. A five analyzers bench allowed us to
collect simultaneously spectra at five different values of the
exchanged wave-vector $Q$ for each single scan. The sample
consisted of a gallium molten droplet kept in sandwich between
two sapphire windows (thickness 0.25 mm). The sample thickness,
($\approx 100 \mu m$), was chosen in order to match the absorption
length at the incident energy of 17794 eV, corresponding to the
(9 9 9) reflection from the silicon analyzers that we utilized
($\delta E \approx 3.0$ meV) \cite{ixs}.

\begin{figure} [h]
\centering
\includegraphics[width=.6\textwidth]{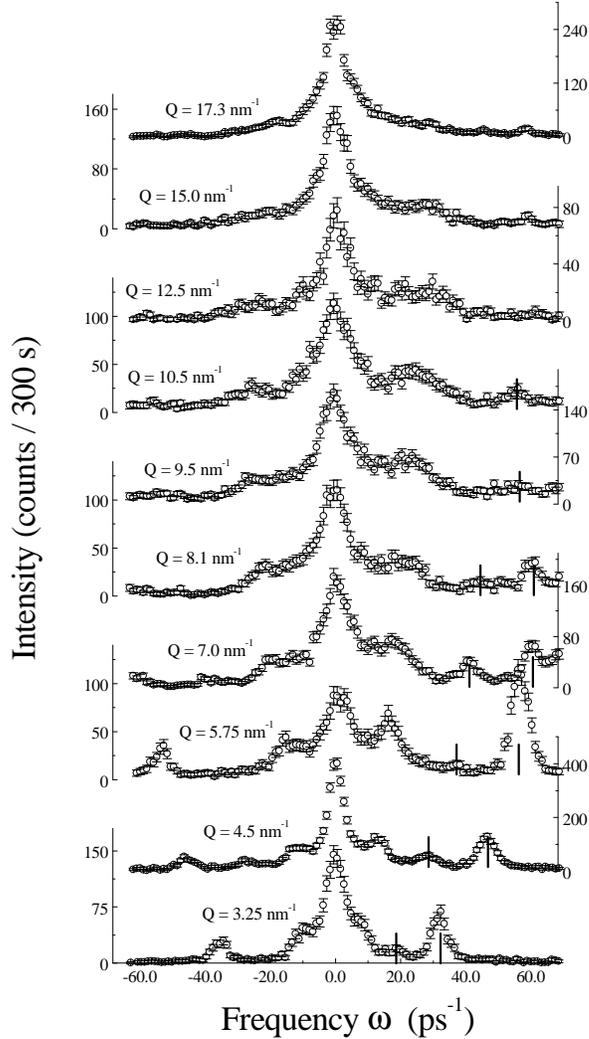}
\vspace{-.01cm}\caption{IXS spectra of liquid Ga ($T=315$ K) at
the indicated $Q$ values. The vertical bars indicate the (Stokes
side) energy positions of the L- and T-acoustic phonons of
sapphire.} \label{panel}
\end{figure}

In Fig. \ref{panel} we report the measured IXS intensity for each
investigated (fixed) $Q$-value. The presence of an acoustic
propagating mode clearly appears from the raw data. These modes
were not observed in Ref. \cite{ber1}, probably due to the limited
accessible cinematic region and to the incoherent scattering
contribution to the spectra. The spurious signal coming from the
longitudinal and transverse phonons of the sapphire windows, due
to the high value of their speed of sound, does not significantly
overlap with the Gallium spectra. In order to extract
quantitative information on the reported excitations, a data
analysis has been performed following a memory function approach
that has been shown to be well footed on several simple fluids
\cite{books,scopixs}. Within the generalized Langevin equation
formalism, it is possible to express the \textit{classical}
dynamic structure factor in terms of a complex memory function,
$M(Q,t)$, related to the interaction details. In particular, in
terms of the real and imaginary part of the Fourier-Laplace
transform of $M(Q,t)$, it holds \cite{books}:

\begin{eqnarray}
S(Q,\omega )= \frac{S(Q)\pi^{-1}\omega _0^2(Q)\tilde{M}^{\prime
}(Q,\omega )}{%
\left[ \omega ^2-\omega _0^2(Q)+\omega \tilde{M}^{\prime \prime
}(Q,\omega )\right] ^2+\left[ \omega \tilde{M}^{\prime }(Q,\omega
)\right] ^2}. \nonumber
\end{eqnarray}

\noindent Here the quantity $\omega _0^2(Q)=KTQ^2/mS(Q)$ is
related to the generalized isothermal sound speed through the
relation $c_t(Q)=\omega_0(Q)/Q$, and can be calculated from the
liquid structure once $S(Q)$ is known. In order to be used as a
test function to fit the experimental data, the above expression
has to be modified to satisfy the detailed balance condition and
must be convoluted with the instrumental resolution function
$R(\omega)$:

\begin{eqnarray}
I^{th}_N(Q,\omega)=\int\frac{\hbar \omega' / KT }{1-e^{-\hbar
\omega' / KT }}S(Q,\omega' )R(\omega- \omega')d\omega '. \nonumber
\label{fitfunction}
\end{eqnarray}

\noindent Taking advantage of the results obtained in several
other liquid metals (Li, Al, Na) \cite{scopixs,scopsim} and fluids
(noble gases) \cite{har,He,Ne}, we utilized a memory function
composed by two relevant time-scales associated to two processes
of viscous origin, plus an exponential contribution that accounts
for the thermal relaxation process. For the two viscous terms we
adopted a simple Debye law (exponential decay) to account for the
structural relaxation and an instantaneous approximation
(delta-function) for the faster, microscopic contribution.
Consequently, the total memory function reads

\begin{eqnarray}
M(Q,t) &=&\left( \gamma -1\right) \omega _{0}^{2}(Q)e^{-D_{T}
Q^{2}t}  \nonumber \label{mem} \\ &+&\Delta_\alpha
^{2}(Q)e^{-t/\tau _{\alpha }(Q)}+2\Gamma_\mu (Q) \delta(t).
\nonumber
\end{eqnarray}

\begin{figure} [h]
\centering
\includegraphics[width=.6\textwidth]{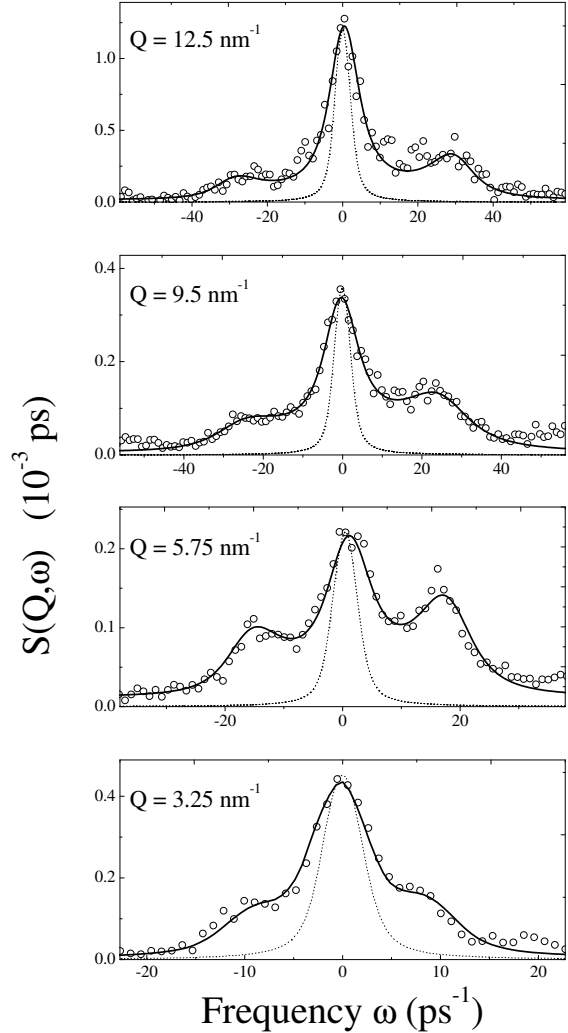}
\caption{Selection of IXS spectra ($\circ$) plotted with the
fitting function (---) described in the text. The instrument
resolution function ($\delta E \approx 3.0$ meV) is also shown
($\cdot \cdot \cdot$).} \label{fit}
\end{figure}

\noindent Thus the free fitting parameters are $\tau_\alpha$ (the
structural relaxation time), $\Delta_\alpha (Q)$ (the structural
relaxation strength) and $\Gamma(Q)$ (associate to the
microscopic relaxation, and representing the Brillouin line-width
in the fully relaxed limit). The value of $\omega_0(Q)$ has been
calculated using the $S(Q)$ data reported in \cite{bf}, while the
specific heat ratio $\gamma$ and thermal diffusivity $D_T$ are
deduced by macroscopic data (their $Q$-dependence have been
neglected). The outcome of the fitting procedure for selected
$Q$-values, i.e. a comparison between the best fitting line-shape
and the experimental spectra, is reported in Fig. \ref{fit}.
Intensities have been normalized using the $S(Q)$ values of Ref.
\cite{bf}.

\begin{figure} [h]
\centering
\includegraphics[width=.6\textwidth]{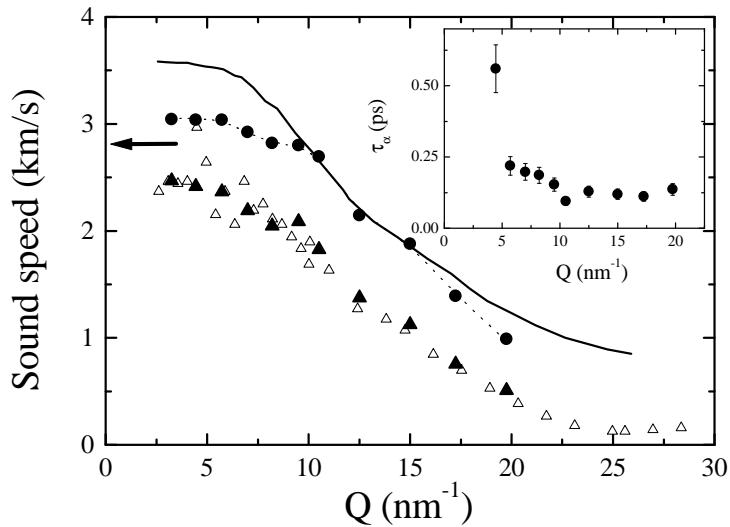}
\vspace{-7.5cm} \caption{Speed of sound deduced from the experimental data
($\bullet$), together with the calculated) $c_\infty (Q)$ (---) and $c_0(Q)$
($\vartriangle$) values \cite{ber1,bf}. The $c_0(Q)$ determined by the fit
($\blacktriangle$), as far as the $c_{\infty \alpha}(Q)$ defined in the text
($\cdot \cdot \cdot$) are also shown.  In the inset we report the values of
the $\alpha$ -relaxation time.} \label{disp}
\end{figure}

In Fig. \ref{disp} we report the apparent sound velocity
\cite{nota} (full dots). At the lower accessible wavevectors, the
measured sound velocity ($\approx$3000 m/s) still exceed the
isothermal value ($\approx$2800 m/s) as deduced by ultrasounds
measurement \cite{ultrasound}, a behaviour (the so called positive
dispersion of the sound velocity) that is common with many other
simple fluids. In the same figure we report the generalized
(Q-dependent) isothermal sound velocity ($\omega_0(Q)/Q$ as
deduced by the $S(Q)$ \cite{bf}) (open triangles) that
constitutes the low frequency limit of the sound speed, as well
as the high frequency limit, $c_\infty (Q) $ (full line)
numerically estimated through the structural data (pair
distribution function, interaction potential) \cite{ber1}. The
full triangles come from the best fitted values of $\omega_0 (Q)$
that -in a separate check fitting session- has been left as free
parameters. The coincidence of the fitting-derived $\omega_0(Q)$
values with those derived from the $S(Q)$ data indicates the
robustness of the fitting model. In the inset the $Q$-dependence
of the structural relaxation time as derived by the fit is
reported. Apart from the very high value found at the lower $Q$
point, mainly because the width of the central line becomes
comparable to the instrument resolution, we find a slightly
decreasing value in the range $0.25>\tau_\alpha >0.17$. It is
worth to point out that the relation $\omega (Q) \tau_\alpha
(Q)>1$ holds in the whole explored $Q$ range, i.e. the structural
relaxation is frozen over the probed timescale. Therefore, the
apparent sound speed coincides with the $c_{\infty \alpha}(Q)
=\sqrt{\omega_0^2(Q)+ \Delta^2_\alpha} /Q$ (dotted line in Fig.
\ref{disp}). The gap observed at low $Q$ between the apparent
speed of sound and the total $c_\infty (Q)$ is due to the
non-negligible strength of the microscopic process, whose
time-scale is much shorter than the inverse of the current
correlation maxima $\omega_l(Q)$, a condition that prevents the
system from reaching the fully unrelaxed regime.

\begin{figure} [h]
\centering
\includegraphics[width=.7\textwidth]{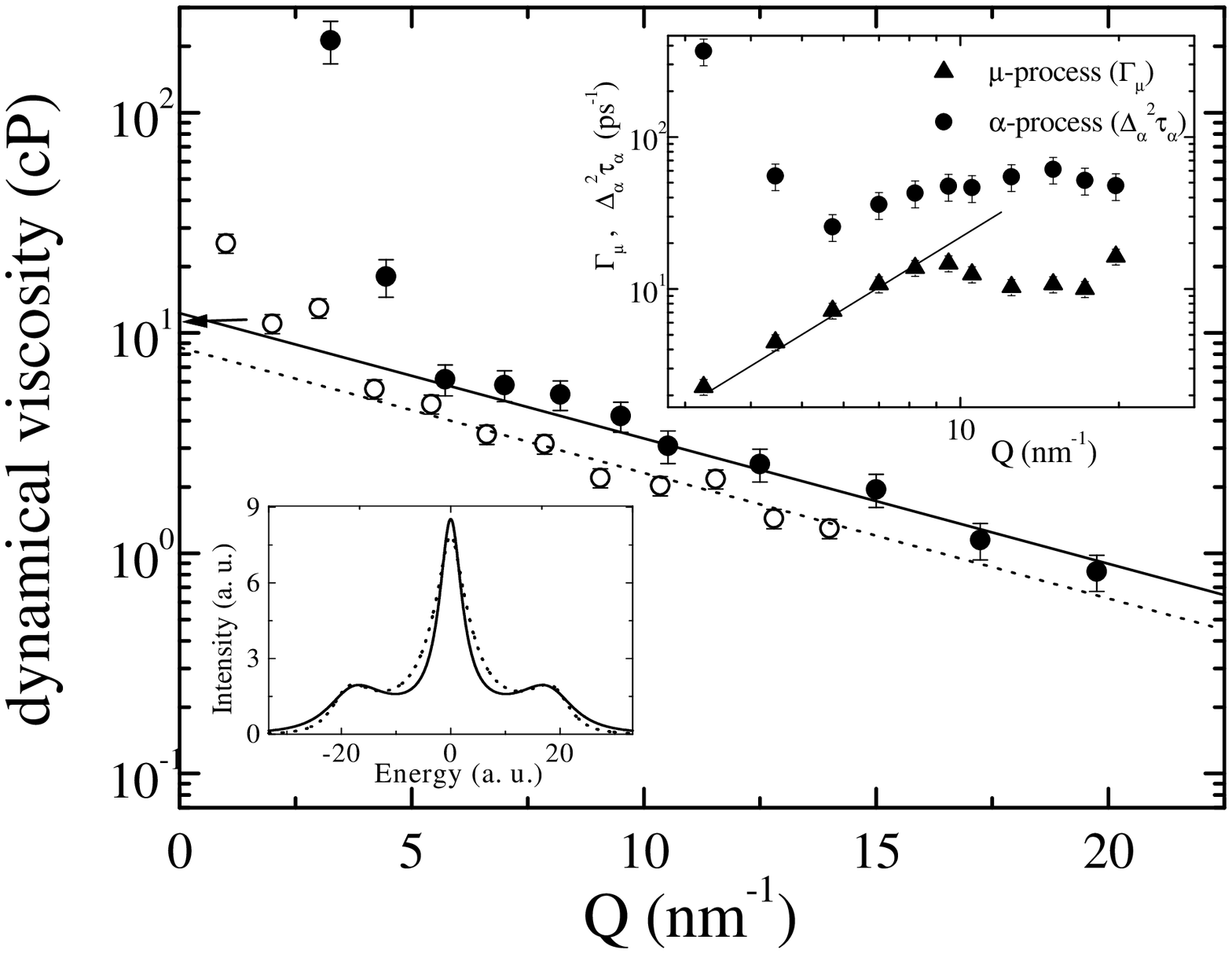}
\vspace{-7.7cm} \caption{Values of $\eta_L(Q)$ as determined by the memory
function parameters for Ga ($\bullet$) and Al ($\circ$) \cite{scopixs}. The
hydrodynamic value for Ga \cite{ber1} is also reported ($\leftarrow$). Upper
inset: partial contributions due to the $\alpha$- ($\bullet$) and $\mu$-
($\blacktriangle$) process, respectively. The full line emphasizes a
noteworthy $Q^2$ behaviour. Lower inset: $S(Q,\omega)$ of Ga (---) and Al
($\cdot \cdot \cdot $) reported in arbitrary units for comparison.}
\label{visco}
\end{figure}

A further relevant quantity that can be extracted from the
fitting parameters is the generalized ($Q$-dependent)
longitudinal viscosity $\eta_L(Q)$, i.e., in the Langevin equation
formalism, the total area of the memory function : $\eta_L(Q)=\rho
(\Delta_\alpha ^2 \tau_\alpha + \Gamma_\mu (Q))/Q^2 $, reported
in Fig. \ref{visco}. In the upper inset we show the two individual
contributions $\Gamma(Q)$ and $\Delta_\alpha ^2(Q)\tau_\alpha(Q)$:
the viscosity associated with the structural rearrangement turns
out to be dominant all over the entire explored $Q$ range; the
contribution coming from the microscopic process (the only one
affecting the Brillouin line-width in the present regime) follows
-at low $Q$- a $Q^2$ behavior as already reported in other
liquids and glasses \cite{scopixs,scopsim,har}. Apart from a low
$Q$ increase, an artifact due to the finite resolution effect and
already observed in other systems \cite{scopixs}, the total
$\eta_L(Q)$ follows an exponential $Q$-dependence, whose origin
deserves further investigations. However, the low $Q$ limit of
$\eta_L(Q)$ can be obtained by extrapolation, and turns out to be
in very good agreement with the value $\eta=11.4$ cP, determined
from the experimental value of the shear viscosity and from the
shear-to-bulk viscosity ratio deduced by structural parameters
\cite{ber1}. In the same figure is also reported the $\eta_L(Q)$
for liquid aluminum at $T_m$ as obtained from previous experiments
\cite{scopixs}, which shows a behaviour similar to that of
Gallium. To further emphasize the similarity of these different
systems, we report in the lower inset a direct spectral
comparison at $Q \approx$9.5 nm$^{-1}$ (the position of the main
peak of the static structure factor is almost the same for Al and
Ga). As the absolute temperatures and the instrumental resolutions
are different in the two cases, we present the best fitted
line-shape according to the model previously discussed. The
energy axes has been arbitrary shrunk to wipe out the effect of
the different sound speed values. Beside obvious quantitative
differences associated -for example- with the different atomic
masses and interactions, the two systems at the melting point
present very close line-shape. This furtherly confirms the
similarity of the high frequency dynamics of liquid Gallium with
that of other simple liquids.

In conclusion, we presented an experimental study of the
collective high frequency dynamics in liquid gallium at the
melting temperature. Evidence for collective acoustic modes has
been found in a $Q$ region extending beyond the hydrodynamic
regime up to one half of the structure factor main peak. A
generalized hydrodynamic analysis allows a quantitative
determination of relevant parameters such as the structural
relaxation time and the generalized viscosities. More
importantly, it reveals how the main features of the collective
dynamics in this system are very similar to the ones reported in
different elements such as Li, Na, Al \cite{scopixs} and noble
fluids \cite{He,Ne}. This is an important indication of how
-beside {\it quantitative} differences- the high frequency
dynamics in simple fluids exhibit universal features which go
beyond system dependent details such as the electronic structure,
bond nature, atomic interaction and structural properties. Since
on the observed time-scale the structure of the liquid is frozen
($\omega_l (Q)\tau_\alpha (Q)>>1$), one can think of the high
frequency dynamics as that of a system with well defined
equilibrium position ("glass"). Therefore, the details of the
dynamics (microscopic relaxation times, residual viscosity) are
fully determined by the vibrations of the disordered structure,
explaining the observed universality \cite{scopsim,har}. Finally,
although we did not find any evidence for additional modes in the
explored $Q$ range, on the basis of the findings of Ref.
\cite{ber2} we believe that further investigations should be
devoted to the higher $Q$ region.

We are thankful to the ESRF staff for the assistance during the
experiment. T.S. gratefully acknowledge M.C. Bellissent Funel for
providing S(Q) data.


\begin{references}
\bibitem{nota} It is worth to recall that the apparent sound
velocity is defined as the ratio of the excitation frequency
$\omega_l(Q)$ to the $Q$ values. The quantity $\omega_l(Q)$, in
turn, is the position of the maximum of the longitudinal current
correlation spectrum $J^{L}(Q,\omega)=\omega^2/Q^2S(Q,\omega)$,
where $S(Q,\omega)$ is obtained by the best fitted parameters,
and therefore is resolution deconvoluted.
\end{references}
\end{document}